\documentclass[iop]{emulateapj}
\usepackage{graphicx,amsmath,amssymb}
\usepackage{natbib}
\usepackage{apjfonts}
\bibpunct[, ]{(}{)}{;}{a}{}{,}

\begin{document}

   \title{Dust extinction bias in the column density distribution of gamma-ray
          bursts; high column density, low redshift GRBs are more
          heavily obscured}

   \author{Darach~Watson}\affil{Dark Cosmology Centre, Niels Bohr Institute, University of Copenhagen, Juliane Maries Vej 30, DK-2100 Copenhagen \O, Denmark;\\\texttt{darach@dark-cosmology.dk}}
\and
   \author{P\'all~Jakobsson}\affil{Centre for Astrophysics and Cosmology, Science Institute, University of Iceland, Dunhaga 5, IS-107 Reykjavik, Iceland;\\\texttt{pja@raunvis.hi.is}}

   \begin{abstract}

The afterglows of gamma-ray bursts (GRBs) have more soft X-ray absorption
than expected from the foreground gas column in the Galaxy.  While the
redshift of the absorption can in general not be constrained from current
X-ray observations, it has been assumed that the absorption is due to metals
in the host galaxy of the GRB.  The large sample of X-ray afterglows and
redshifts now available allows the construction of statistically meaningful
distributions of the metal column densities.  We construct such a sample and
show, as found in previous studies, that the typical absorbing column
density ($N_{\rm H_X}$) increases substantially with redshift, with few high
column density objects found at low to moderate redshifts.  We show,
however, that when highly extinguished bursts are included in the sample,
using redshifts from their host galaxies, high column density sources are
also found at low to moderate redshift.  We infer
from individual objects in the sample and from observations of blazars, that
the increase in column density with redshift is unlikely to be related to
metals in the intergalactic medium or intervening absorbers.  Instead we
show that the origin of the apparent increase with redshift is primarily due
to dust extinction bias: GRBs with high X-ray absorption column densities
found at $z\lesssim4$ typically have very high dust extinction column
densities, while those found at the highest redshifts do not.  It is unclear
how such a strongly evolving $N_{\rm H_X}/A_V$ ratio would arise, and based
on current data, remains a puzzle.

   \end{abstract}

   \maketitle

%
%
\section{Introduction\label{introduction}}

While it is now generally accepted that long-duration gamma-ray bursts
(GRBs) primarily originate in the explosions of massive stars due to their
association with type Ic supernovae, the precise nature of the progenitors
and the environment in which the burst occurs are not known.  Most progress
to date has been made through afterglow observations; providing redshifts,
emission mechanisms and information on the host galaxies.  However the X-ray
afterglows are still poorly understood.  Indeed, one of the outstanding
puzzles in understanding long GRBs is the nature and origin of the soft
X-ray absorption observed in the majority of afterglows.  Most GRB
afterglows show evidence of absorption in the soft end of the X-ray spectrum
significantly in excess of what is expected from the Galactic gas column. 
This has been known statistically from samples since the \emph{BeppoSAX} era
\citep{2001ApJ...549L.209G}, though first observed at high confidence in a
single spectrum with XMM-\emph{Newton} \citep{2002A&A...395L..41W}.  It has
generally been assumed from the beginning that the soft X-ray opacity is due
to photoelectric absorption by the inner shells of the atoms in a column of
metals -- primarily O, Si, S, Fe, He -- in the host galaxy of the GRB, in a
fashion directly comparable to the X-ray absorption observed due to gas in
the Galaxy.  However, the absorption was quickly realised not to be directly
analogous to Galactic soft X-ray absorption.  The X-ray absorption in the
Galaxy is strongly correlated with the dust and \ion{H}{1} column densities
\citep[see][and references therein]{2011A&A...533A..16W}.  However, for GRBs,
the correlation with dust extinction was not clear, and if it existed, was
certainly at least an order of magnitude lower in dust-to-metals ratio
compared to the local group \citep{2011A&A...532A.143Z,2010MNRAS.401.2773S}. 
There was also no obvious correlation with the \ion{H}{1} column densities
\citep{2007ApJ...660L.101W,2010MNRAS.402.2429C,2011A&A...525A.113S}.  More
recently a further puzzle was added.  \citet{2010MNRAS.402.2429C} showed
that the observed X-ray absorptions rose with redshift, with the highest
column density objects ($\log{N_{\rm H_X}}\sim23$) appearing at the highest
redshifts, and no comparably high column densities occurring at low
redshifts ($\log{N_{\rm H_X}}\lesssim22$ at $z<1.5$).  This result was
particularly puzzling since the X-ray absorption measures the total metal
column density, and the gas metallicity is expected to decrease rather than
increase to high redshift.

It was noted by \citet{2011ApJ...734...26B} that the \emph{observed} opacity
at low energies, while high at low redshift, tended toward an asymptotic
value at $z\gtrsim2$.  This was interpreted as possible evidence for the
detection of absorption by a diffuse, highly-ionised intergalactic medium
\citep{2011ApJ...734...26B}.  Such an interpretation has the virtue that it
would solve the problems of the lack of correlation observed between the
Ly$\alpha$-determined \ion{H}{1} column densities and the X-ray column
densities in GRB afterglows, and the very low apparent dust-to-metals ratios. 

Finally, a recent investigation of a largely redshift-complete sample of
bright GRBs \citep{2012MNRAS.421.1697C} found a statistically insignificant
mild increase of X-ray absorption with redshift and interpreted it as due to
increasing absorption by intervening systems in higher redshift GRBs.

In this paper we address the nature of the X-ray absorption in GRB
afterglows; we investigate the apparently increasing
absorption with redshift, the claim of a possible detection of the warm-hot
intergalactic medium, and the role of dust extinction.  In
section~\ref{observations} we detail the data used and our data analysis
method.  In section~\ref{results} we provide the results of our analysis. 
Section~\ref{discussion} contains a discussion on the interpretation of
these results.  

%
%
\section{Observational data and methods}\label{observations} 

We analysed the XRT data from every long-duration burst observed by \emph{Swift} up to
November 2010.  For each GRB in this set with a known redshift, we used the
spectra produced by the auto-analysis of \citet{2009MNRAS.397.1177E} with
the corresponding response files and fit a model consisting of a power-law
absorbed by Galactic gas and gas at the redshift of the GRB to each dataset. 
We obtained redshifts for all bursts from the literature, primarily from
\citet{2009ApJS..185..526F}, \citet{2012arXiv1205.3490J},
\citet{2012arXiv1205.4036K}, and GCNs.  This resulted in 175 GRBs. The data from windowed timing (WT)
and photon counting (PC) modes were fit separately.  The Galactic gas was
modelled with an absorber fixed at a level set by the dust extinction in the
direction of the GRB \citep{1998ApJ...500..525S} 
with $N_{\rm H_{MW}} = 2.2\times10^{21}A_V$, as
suggested by \citet{2011A&A...533A..16W}.  The method used to determine the
Galactic column density of metals (whether using the neutral hydrogen or the
dust as a tracer), does not appear to significantly affect the results as we
obtain similar values for the excess absorption as previous authors where
the GRBs analysed are in common \citep{2010MNRAS.402.2429C}.  The absorption
at the redshift of the GRB was allowed free to vary.  The model used was
\texttt{tbabs(ztbabs(pow))} in Xspec with metallicities from
\citet{1989GeCoA..53..197A}.  We used the absorption model of
\citet{2000ApJ...542..914W} to fit the data as the atomic cross-sections are
more accurate.  We use the metallicity of \citet{1989GeCoA..53..197A} for
ease of comparison with previous results, which generally use these
abundances.  It should be noted that while these abundances are
significantly higher than the best estimates of the solar photosphere
abundances \citep{2009ARA&A..47..481A}, they are likely a better estimate of
the typical Galactic ISM abundance \citep{2011A&A...533A..16W}.  In any
event, as noted above, we sidestep this metallicity conversion problem for
the Galactic absorption simply by using the measured relationship between
dust and X-ray absorption.  However, it should be born in mind that the
excess equivalent hydrogen column densities we report here are determined
assuming an abundance approximately 50\% higher than solar.  Thus, for
almost any GRB host these numbers are lower limits to the actual gas column
density and if the real gas column is sought should be corrected for the
probable metallicity of the GRB host galaxy.  In this paper, we simply use
the equivalent hydrogen column density as a proxy for the total metal column
density.  We do this for comparison with previous work, since this is what
has been done by many authors before; and because we cannot determine
individual metal column densities, the hydrogen proxy is the easiest one to
deal with in a simple way.

The absorption in the X-ray afterglows of some GRBs appears to decrease as a
function of time
\citep[e.g.][]{2005A&A...442L..21S,2007A&A...462..565G,2007ApJ...654L..17C}. 
For this reason, where the WT and PC data gave statistically different
values of the absorption, the later PC data gives us a conservative (low)
value of the absorption.  However, the WT data often has considerably higher
signal.  Therefore, where the PC and WT mode data gave results consistent
within $1\sigma$ (68\% confidence), the value with the smallest uncertainty
was used.  Where the results were discrepant at $>1\sigma$, the PC value was
used.  This procedure will be conservative in the sense that it will tend to
lower values of $N_{\rm H_X}$.

We determined extinction estimates for the GRB afterglows from the works of
\citet{2011A&A...532A.143Z}, \citet{2011A&A...526A..30G},
\citet{2010MNRAS.401.2773S}, and \citet{2010ApJ...720.1513K} primarily.  In
cases where no explicit estimate of extinction could be found, we used the
deepest limits from optical/NIR observations from the literature and the
corresponding X-ray data, to determine values or limits on $\beta_{\rm XOX}$
\citep{2009ApJ...699.1087V,2004ApJ...617L..21J}, where $\beta_{\rm XOX} =
\beta_{\rm X} - \beta_{\rm OX}$.  From these data, we could
then also derive limits on the restframe extinction: we use the theoretical
and empirical determination that $\Delta\beta=0.5$
\citep{2011A&A...532A.143Z,1998ApJ...497L..17S}, and that hence, $\beta_{\rm
XOX}\le0.5$.  We then assumed an SMC extinction curve, found by all works to
date to be most typical of the extinction for most GRB afterglows
\citep{2011A&A...532A.143Z,2011A&A...526A..30G,2010MNRAS.401.2773S,2010ApJ...720.1513K},
to determine a limit on the minimum extinction required to make the
optical/NIR photometry consistent with $\beta_{\rm XOX}\leq0.5$.

\begin{figure}
 \includegraphics[angle=-90,bb=70 126 551 655,clip=,width=\columnwidth]{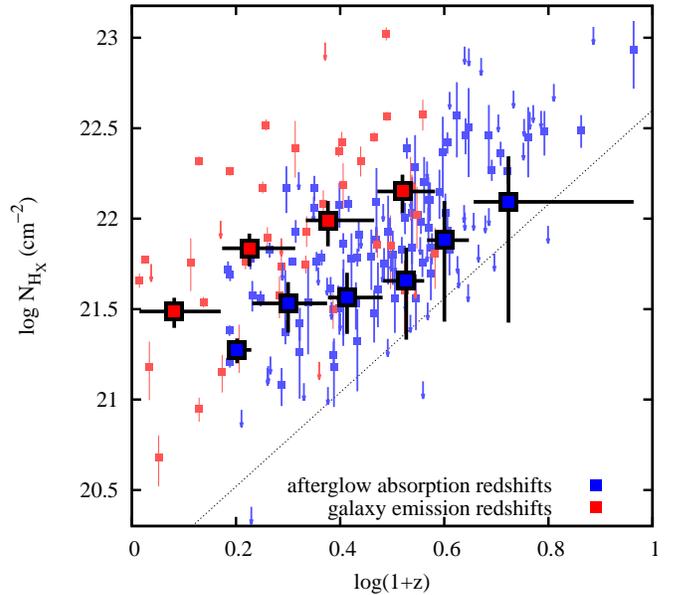}
 \caption{The X-ray absorption of GRB afterglows as a function of redshift. 
          Overplotted in larger symbols is the mean absorption in a given
          redshift window taking into account upper limits.  GRBs with
          redshifts obtained from absorption lines in the optical afterglow
          are plotted in blue.  Those with redshifts obtained from host
          galaxy emission are plotted in red.  The bias introduced due to
          dust extinction is clear in the systematically higher column
          densities in the emission redshift GRBs ($\Delta\log{N_{\rm
          H_X}}=0.31\pm0.08$).  The host emission
          redshift sample still has lower absorption at low redshifts possibly
          due to significant remaining incompleteness in the sample, and
          there is no evidence for a more constant distribution
          of column densities than in the afterglow-selected sample.
          The
          dotted line marks a $10^{20}$\,cm$^{-2}$ absorber at $z=0$ evolved
          as $(1+z)^{2.5}$, showing the approximate detectability threshold
          for absorption as a function of redshift for a typical
          \emph{Swift}-XRT GRB afterglow.}
 \label{fig:mean_nh}
\end{figure}

\begin{figure}
 \includegraphics[angle=-90,bb=69 132 548 646,clip=,width=\columnwidth]{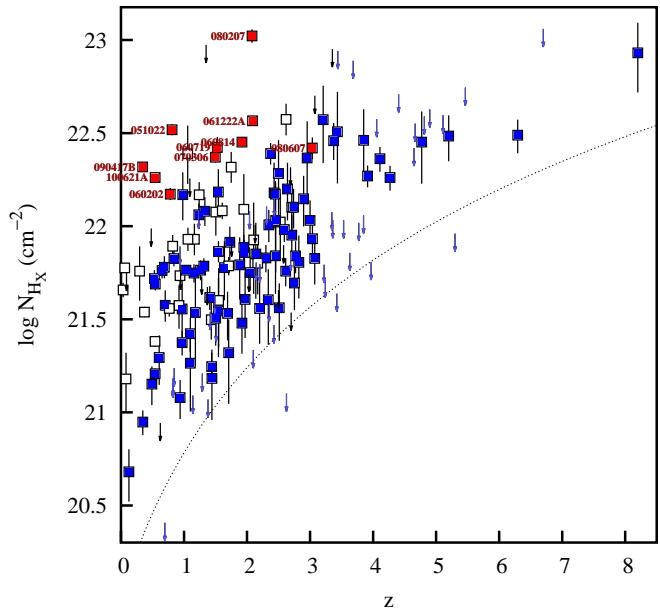}
 \caption{X-ray absorption of GRB afterglows as a function of redshift.
          Afterglows with extinction $A_V>1.5$ are plotted in red, those
          with $A_V<1.5$ in blue.  Afterglows with unknown extinction or
          with 95\% uncertainties spanning this boundary are plotted with
          open squares.  The dotted line is as explained in
          Fig.~\ref{fig:mean_nh}.  The X-ray absorptions rise systematically
          with redshift in the absence of the high extinction afterglows,
          similar to previous findings without these high extinction events. 
          The upper bound of the absorption appears roughly constant with
          redshift when high extinction events are included.}
 \label{fig:z_nh}
\end{figure}

%
%
\section{Results}\label{results}

As with previous work, we find very significant absorbing column densities
in excess of the Galactic value for most bursts.  Similarly, we also find
that the mean absorption increases with redshift (Fig.~\ref{fig:mean_nh}). 
And while we know that the lower bound for detection of absorption increases
strongly as a function of redshift, we do not reproduce the most interesting
previous finding, that the upper envelope of the absorbing column density
increases with redshift \citep{2010MNRAS.402.2429C}.  In the distribution
shown in Figs.~\ref{fig:mean_nh} and \ref{fig:z_nh}, the vast majority of
objects have $\log{N_{\rm H_X}}\lesssim22.6$, with only two outliers above
this value: one at $z\sim2.2$, and the other at $z\sim8.2$.  We do not have
the total absence of GRBs with $\log{N_{\rm H_X}}\gtrsim22$ at $z\lesssim2$
found previously \citep{2010MNRAS.402.2429C,2011ApJ...734...26B}.  The more
complete, bright sample analysed in \citet{2012MNRAS.421.1697C} showed this
effect as well, with a few high column density objects appearing at low redshift,
allowing them to infer that the previous absence of such bursts was a selection bias.

So, why were these low redshift, high absorption GRBs missing from previous
analyses?  Coding the distribution by dust extinction, the answer is
immediately apparent (Fig.~\ref{fig:z_nh}); almost all of the low redshift,
high absorption GRBs have high extinction ($A_V>1.5$).  These GRBs are: 051022,
060202, 060719, 060814, 061222A, 070306, 070521, 080207, 080607, and 090417B. The high restframe
extinction of these objects makes it very difficult to obtain a redshift
from the optical afterglow \citep[see, for example][who examine eight
highly-extinguished objects and find five at low redshift with high X-ray
absorbing column densities]{2011A&A...534A.108K}.  At $z=1.5$, $A_V=1.5$ is
equivalent to a factor of more than 50 suppression of the observed $V$-band
flux.  The vast majority of these high extinction objects have redshifts
obtained from the host galaxy.  A good example is GRB\,080207 with a
redshift estimated to be $z\sim2$ from photometric observations of the host
\citep{2011ApJ...736L..36H,2011arXiv1109.3167S}.  Its afterglow extinction
is estimated to be $A_V\gtrsim3.5$.  This corresponds to an extinction of
the flux by a factor of $\gtrsim10^5$ in the \emph{observed} $V$-band.

Excluding these high extinction objects, we retrieve the rising upper
envelope observed by \citet{2010MNRAS.402.2429C}.  Indeed, it was noted by
\citet{2010MNRAS.402.2429C} that dust extinction bias might explain the
rising upper envelope they observed, though the proposed explanation related
to flatter extinction curves at higher redshifts is unlikely to be correct
since even relatively flat extinction curves have curvature
\citep[e.g.][]{2008MNRAS.384.1725H,2004Natur.431..533M}, and with such high
$A_V$s would either render the GRB optical afterglows undetectable, or would
be clearly noticeable in the broadband SED fitting
\citep{2011ApJ...735....2Z}.  On the other hand, including all the available
data, it seems clear that the upper bound of the $N_{\rm H_X}$ distribution
is the same at low and high redshifts.

In spite of the similarity of the upper bound at high and low redshifts, it
is still apparent that even in our more complete sample, the fraction of
objects with a very high column density at high redshifts is still very
large compared to lower redshift.  This can be seen from the high ratio of
detections to upper limits at high redshift, which, if the distribution was
the same at all redshifts, would be significantly lower, or, equivalently,
from the mean column density which still increases with redshift
(Fig.~\ref{fig:mean_nh}).  However our sample is not complete and almost
certainly still has a strong bias against highly obscured objects. 
Therefore, it is not clear whether the column density distribution is the
same at low and high redshifts.  If it is, then a significant majority of
GRBs without redshifts must be at low redshift \citep[see, among others,][]{2009ApJS..185..526F,2011A&A...534A.108K},
which means that previous estimates of the mean redshifts of GRBs were
skewed to excessively high values \citep[e.g.][]{2006A&A...447..897J}. 
Given the data so far, of course they would also have to be highly
extinguished.  The recent, largely complete, sample of bright GRBs analysed
for their X-ray absorptions by \citet{2012MNRAS.421.1697C} still shows a
mild increase in X-ray absorption with redshift, which they attribute to
foreground absorbers.

%
%

\section{What is the origin of the X-ray absorption in GRB afterglows?}\label{discussion}

The results outlined in the previous section raise more questions than they
answer.  The peculiarity now is no longer why there are no high absorption
GRBs at low redshift, but why do high-absorption, low-extinction objects
show up at high redshift, but not at lower redshifts?  While we might expect
a strong bias against getting redshifts for dust-obscured GRBs at high
redshift, we would not expect any bias against low-extinction GRBs at low
redshift.

\subsection{Intrinsic curvature}

We can readily exclude intrinsic curvature as an explanation for the soft
X-ray downturn in the general case because the shape of the downturn does
not fit typical GRB models, requiring low energy slopes different from those
observed in the prompt phase \citep{2008ApJ...677.1168K}
because the measured absorption is occasionally found to be constant in
spite of large spectral changes, for example in GRB\,100901A where we find
the spectral slope changes from $\Gamma=1.7\pm0.03$ to $2.2\pm0.05$ between
the WT and PC data, but the excess absorptions are
$4.1^{+0.4}_{-0.3}\times10^{21}$\,cm$^{-2}$ and
$4.1\pm0.6\times10^{21}$\,cm$^{-2}$.  As an aside, intrinsic
spectral curvature is observed in many blazars, but their spectra are
considerably more complex than GRB afterglows and the difference in their
low energy slopes is typically small, with $\Delta\beta\lesssim0.5$
\citep{2005A&A...433.1163D,2005ApJ...625..727P}.  Indeed, it should be noted
that intrinsic curvature has never been invoked as a general explanation,
though it may play a role in the few objects at the highest redshifts
\citep{2007ApJ...663..407B}.

\subsection{The warm-hot intergalactic medium}

We can also exclude a smooth, highly ionised intergalactic medium, the
so-called warm-hot intergalactic medium (WHIM), as the explanation for the
absorption, as proposed by \citet{2011ApJ...734...26B}.  We can see very
quickly that there are many GRB afterglows at $z\gtrsim3$ with absorptions
well below the apparent proposed level of the smooth WHIM
(Fig.~\ref{fig:z_nh}).

A more general argument is a structured WHIM, where it is only the average
opacity above $z\sim2$ that would tend to a detectable value while
individual sightlines could have disparate values.  As opposed to the smooth
WHIM, individual GRB afterglows with low absorption values could be
reconciled to this model.  A potentially useful sample to compare to are
type\,1 AGN.  Their spectra are known in most cases to be free of gas or
dust absorption, and to show no evolution in absorbing column density as a
function of redshift up to $z\sim3$ \citep{2010A&A...510A..35M}.

However, an analysis of the X-ray spectra of small samples of high-redshift
AGN observed with \emph{XMM-Newton} appear to show substantial absorbing
column densities in the radio-loud, but not the radio-quiet AGN
\citep{2005MNRAS.364..195P,2006MNRAS.368..985Y,2011ApJ...738...53S}.

In the high-redshift, radio-loud AGN there are are approximately 50\% which
show a downturn at low energies, and about 25\% that show a noticeable
upturn, suggesting that the spectra are not simple power-laws, and may be
considerably more complex \citep{2005MNRAS.364..195P}.  None of the seven
radio-quiet objects in the \citet{2005MNRAS.364..195P} sample show a
significant up- or downturn.  \citet{2007ApJ...665..980T} have suggested
that the downturn observed in at least some, and possibly all
\citep{2007ApJ...669..884S}, radio-loud AGN is intrinsic curvature of the
low energy side of the inverse Compton emission component, and successfully
modelled this in the object RBS\,315.  Indeed, intrinsic
curvature is known in the spectra of blazars at low redshift. It mimics absorption and
is present up to fairly hard energies
\citep{2005ApJ...625..727P,2000ApJ...541..166F}.
Furthermore, the apparent `absorption' observed in low redshift blazars is
peculiar in that no atomic absorption edges or lines are ever observed
\citep[e.g.][]{2004A&A...418..459W,2004A&A...417...61B}, further
strengthening the conclusion that the observed downturns are in fact due to
intrinsic curvature.  Therefore, finding evidence of such curvature in some
higher redshift radio-loud AGN should not be a surprise.  In the analysis of
RBS\,315, \citet{2007ApJ...665..980T} examined two epochs of
\emph{XMM-Newton} spectroscopy taken three years apart.  Fitting both
datasets with an absorbed powerlaw, they found the measured absorption had
increased by more than 50\%.  This apparent variability of the downturn
argues forcibly against a WHIM origin, since the WHIM is unlikely to vary on
a 3 year timescale.

\subsection{High density, low-ionisation foreground absorbers}
It might also be argued that the apparent increase of the X-ray
absorption with redshift is related to high density (neutral or
low-ionisation) intervening absorbers along the line of sight to the GRBs. 
And indeed, the paucity of low column density systems at high redshift is
curious, though currently not highly significant statistically given that
the detectability threshold at high redshift is so high. 
\citet{2012MNRAS.421.1697C} calculate the distribution of Ly$\alpha$
absorbers with redshift based on observational data and show that it is too
low to explain the high absorptions observed at high redshift.  They contend
that the number of absorbers foreground to GRBs may be double that observed
foreground to QSOs based on the numbers of high equivalent width \ion{Mg}{2}
absorbers discovered \citep{2009A&A...503..771V}.  Using this larger number,
they find that the neutral absorbers could plausibly explain the high X-ray
absorption systems found at the highest redshifts.  However, they do not
take into account the metallicity evolution of DLAs with redshift, but
assume that the mean metallicity of the absorbers is solar.  This is
important since the X-ray absorption measures metals not hydrogen. 
\citet{2003ApJ...595L...9P} demonstrate that the metallicity of intervening
absorbers evolves strongly with redshift, such that the contribution from
$z=2-4$ absorbers will be 0.5--1.0\,dex lower than plotted in Fig.\ 2 of
\citet{2012MNRAS.421.1697C}.  This demonstrates that even by doubling the
number of intervening systems, neutral absorbers along the line of sight are
simply insufficient to induce the observed X-ray absorption in GRBs. 
Furthermore, such absorbers should be detected in the optical unless they
are at very low redshift ($z\lesssim0.3$) or have very low column density. 
This seems unlikely on either count, since one would have to pack a large
number of absorbing systems into the redshift space $z<0.3$ and have far
fewer at $z>0.3$, and at $z<0.3$, we would expect to detect such absorbing
systems relatively easily in emission as galaxies close to the lines of
sight.  In addition, the excess of \ion{Mg}{2} absorbers is only found at
very large equivalent widths; at lower equivalent widths the numbers for
GRBs and QSOs are the same \citep{2009A&A...503..771V}.  Thus doubling the
population of low column density sources is unjustified.  Another way of
looking at this is to examine the relationship found by
\citep{2009MNRAS.393..808M} between $N_{\rm HI}$ and \ion{Mg}{2} equivalent
width for QSO intervening Ly$\alpha$ systems.  The mean number of high
equivalent width ($W_{\lambda,0} > 1$\,\AA\ \ion{Mg}{2} absorbers found by
\citet{2009A&A...503..771V} is 0.7 per redshift interval for GRBs.  The
typical equivalent width of these systems is a few \AA.  This corresponds to
a $N_{\rm HI} \sim 1\times10^{20}$ for $W_{\lambda,0} = 2$\,\AA.  Regardless
of the redshift, this is vastly insufficient to explain the X-ray
absorption.  Even assuming the \ion{Mg}{2} absorber is at $z=0.5$, the
apparent contribution to a $z=4$ GRB is only $2\times10^{21}$\,cm$^{-2}$. 
We would then require at least five such $z<1$ systems for every $z=4$ GRB,
or a single system with $W_{\lambda,0} > 6$\,\AA, at $z=0.5$ in most $z>4$
GRBs, which is the largest equivalent width ever observed for any GRB to
date.  In other words, most GRBs would need to have absorption systems
similar to, or larger than, GRB\,991216 \citep{2006A&A...447..145V}, and this
is not observed \citep{2009A&A...503..771V}.  In addition, we would have not
to observe the corresponding galaxy at relatively low impact factor at such
low redshifts.  Finally, this solution does not really answer the
fundamental problem of the peculiarly low dust-to-metals ratio; we would
expect to observe substantial extinction if the absorbers were predominantly
at moderate redshift, since DLAs appear to have a dust-to-metals ratio
similar to the local group \citep{1998ApJ...493..583V,2009MNRAS.393..808M}.

\section{The origin of the apparent increase in X-ray absorbing
            column density with redshift}

We have shown that the apparent increase in X-ray absorbing column densities
with redshift is mitigated when dust-extinguished bursts are included in the
sample, i.e.\ we fill in the high-absorption, low-redshift bursts that are
missing from Fig.~2 of \citet{2010MNRAS.402.2429C}.  We have also shown that
the bursts with the highest absorbing column densities are also typically
the most dust-extinguished at low-to-moderate redshift ($z\lesssim4$). 
This
is similar to the conclusion that there is a tight correlation between `dark' GRBs
and high X-ray absorption GRBs found recently \citep{2012MNRAS.421.1697C}.
Furthermore, we still lack redshifts for many bursts -- many of these are
dark bursts and are almost certainly heavily extinguished.
If most of the most highly-absorber bursts are at low redshift, i.e.\ are similar to
the other obscured bursts we have found, they may provide the population
needed to correct the
apparent increase in X-ray absorption to high redshift. We therefore
conclude that the increasing X-ray absorption observed so far as we go to high
redshift is principally due to a dust-extinction bias.  However, a
complete, unbiased sample of sufficient size is required to answer this
question. \citet{2012MNRAS.421.1697C} argue for a moderate remaining
increase with redshift in their 90\% complete sample of bright GRBs, but
have insufficient statistical power to make the claim with confidence.

\begin{figure}
 \includegraphics[angle=-90,bb=60 105 538 672,clip=,width=\columnwidth]{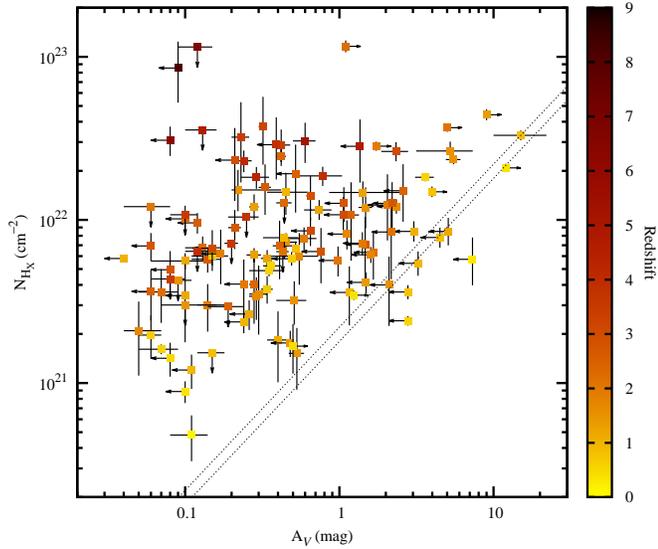}
 \caption{X-ray absorption versus extinction in GRB afterglows. Points are
          coded as a function of redshift. The dashed lines mark the
          approximate limits of the metals-to-dust ratios reported for the
          local group of galaxies.
          The
          absence of significant numbers of low redshift objects with high
          absorption and low extinction is noticeable, i.e.\ there is a
          lack of low redshift objects in the upper left part of the
          plot.}
 \label{fig:nx_av_z}
\end{figure}

An increasing X-ray absorption due to a dust extinction bias is contrary to
what we would na\"{\i}vely expect; extinction at high redshift has a
dramatic effect on the observed optical/NIR flux since we are observing the
restframe far-UV where dust extinction is at its most severe.  At low
redshift we see that the most X-ray absorbed bursts have higher extinction. 
One would expect that bursts at high redshifts with large X-ray column
densities would very rarely be detected.  The opposite seems to be the case. 
This is apparent in Fig.~\ref{fig:nx_av_z}, where there is a clear redshift
gradient in the metals-to-dust ratio plane; in particular, no high
metals-to-dust ratio objects are detected at low redshift.  Such objects
should be easily detected.

The simplest interpretation of this unexpected behaviour is that the
fraction of metals in the dust phase is dramatically lower at $z\gtrsim4$.
However, this does not seem a very tenable hypothesis, since we know that
dust does form very rapidly following formation of metals
\citep{2011Sci...333.1258M,2010ApJ...712..942M}.  Furthermore, there are fairly
strong indications that in many environments as well as in the
low-ionisation phase of the afterglow absorption, the dust-to-metals ratio
is very roughly constant \citep{2011PhDT......deCia,1998ApJ...493..583V,2009ApJ...692..677D}.

If a change in the metals-to-dust ratio were responsible, it would imply
that the formation efficiency of dust out of available metals is at least
one order of magnitude lower at $z\gtrsim4$ than at lower redshifts.  But as
indicated above, we do not find this a likely explanation.  There must be
other explanations for this apparent change, and we note that the mean
metallicity decreases strongly with redshift \citep{2003ApJ...595L...9P} and
this change may reflect the more pristine environments found earlier in the
age of the universe.  It will be difficult to solve this mystery for certain
until we have a clear idea of what causes the X-ray absorption and precisely
where it occurs.

%
%

\begin{acknowledgements} 

The Dark Cosmology Centre is funded by the DNRF. PJ acknowledges support by
a Project Grant from the Icelandic Research Fund.  We would like to thank
Andrew Zirm, Tayyaba Zafar, Nial Tanvir, Adam Miller, Daniele Malesani,
Thomas Kr\"uhler, Jens Hjorth, and Anja C.\ Andersen for discussions and
comments on the manuscript, and Javier Gorosabel for an estimate of $A_V$
for GRB\,090516A.

\end{acknowledgements}


\begin{thebibliography}{55}
\expandafter\ifx\csname natexlab\endcsname\relax\def\natexlab#1{#1}\fi

\bibitem[{{Anders} \& {Grevesse}(1989)}]{1989GeCoA..53..197A}
{Anders}, E., \& {Grevesse}, N. 1989, \gca, 53, 197

\bibitem[{{Asplund} {et~al.}(2009){Asplund}, {Grevesse}, {Sauval}, \&
  {Scott}}]{2009ARA&A..47..481A}
{Asplund}, M., {Grevesse}, N., {Sauval}, A.~J., \& {Scott}, P. 2009, \araa, 47,
  481

\bibitem[{{Behar} {et~al.}(2011){Behar}, {Dado}, {Dar}, \&
  {Laor}}]{2011ApJ...734...26B}
{Behar}, E., {Dado}, S., {Dar}, A., \& {Laor}, A. 2011, \apj, 734, 26

\bibitem[{{Blustin} {et~al.}(2004){Blustin}, {Page}, \&
  {Branduardi-Raymont}}]{2004A&A...417...61B}
{Blustin}, A.~J., {Page}, M.~J., \& {Branduardi-Raymont}, G. 2004, \aap, 417,
  61

\bibitem[{{Butler} \& {Kocevski}(2007)}]{2007ApJ...663..407B}
{Butler}, N.~R., \& {Kocevski}, D. 2007, \apj, 663, 407

\bibitem[{{Campana} {et~al.}(2010){Campana}, {Th{\"o}ne}, {de Ugarte Postigo},
  {Tagliaferri}, {Moretti}, \& {Covino}}]{2010MNRAS.402.2429C}
{Campana}, S., {Th{\"o}ne}, C.~C., {de Ugarte Postigo}, A., {Tagliaferri}, G.,
  {Moretti}, A., \& {Covino}, S. 2010, \mnras, 402, 2429

\bibitem[{{Campana} {et~al.}(2007){Campana}, {Lazzati}, {Ripamonti}, {Perna},
  {Covino}, {Tagliaferri}, {Moretti}, {Romano}, {Cusumano}, \&
  {Chincarini}}]{2007ApJ...654L..17C}
{Campana}, S., {et~al.} 2007, \apj, 654, L17

\bibitem[{{Campana} {et~al.}(2012){Campana}, {Salvaterra}, {Melandri},
  {Vergani}, {Covino}, {D'Avanzo}, {Fugazza}, {Ghisellini}, {Sbarufatti}, \&
  {Tagliaferri}}]{2012MNRAS.421.1697C}
---. 2012, \mnras, 421, 1697

\bibitem[{{Dai} \& {Kochanek}(2009)}]{2009ApJ...692..677D}
{Dai}, X., \& {Kochanek}, C.~S. 2009, \apj, 692, 677

\bibitem[{{de~Cia}(2011)}]{2011PhDT......deCia}
{de~Cia}, A. 2011, PhD thesis, School of Engineering and natural Sciences,
  Faculty of Physical Sciences, University of Iceland, Reykjav\'{\i}k

\bibitem[{{Donato} {et~al.}(2005){Donato}, {Sambruna}, \&
  {Gliozzi}}]{2005A&A...433.1163D}
{Donato}, D., {Sambruna}, R.~M., \& {Gliozzi}, M. 2005, \aap, 433, 1163

\bibitem[{{Evans} {et~al.}(2009){Evans}, {Beardmore}, {Page}, {Osborne},
  {O'Brien}, {Willingale}, {Starling}, {Burrows}, {Godet}, {Vetere}, {Racusin},
  {Goad}, {Wiersema}, {Angelini}, {Capalbi}, {Chincarini}, {Gehrels}, {Kennea},
  {Margutti}, {Morris}, {Mountford}, {Pagani}, {Perri}, {Romano}, \&
  {Tanvir}}]{2009MNRAS.397.1177E}
{Evans}, P.~A., {et~al.} 2009, \mnras, 397, 1177

\bibitem[{{Fossati} {et~al.}(2000){Fossati}, {Celotti}, {Chiaberge}, {Zhang},
  {Chiappetti}, {Ghisellini}, {Maraschi}, {Tavecchio}, {Pian}, \&
  {Treves}}]{2000ApJ...541..166F}
{Fossati}, G., {et~al.} 2000, \apj, 541, 166

\bibitem[{{Fynbo} {et~al.}(2009){Fynbo}, {Jakobsson}, {Prochaska}, {Malesani},
  {Ledoux}, {de Ugarte Postigo}, {Nardini}, {Vreeswijk}, {Wiersema}, {Hjorth},
  {Sollerman}, {Chen}, {Th{\"o}ne}, {Bj{\"o}rnsson}, {Bloom}, {Castro-Tirado},
  {Christensen}, {De Cia}, {Fruchter}, {Gorosabel}, {Graham}, {Jaunsen},
  {Jensen}, {Kann}, {Kouveliotou}, {Levan}, {Maund}, {Masetti},
  {Milvang-Jensen}, {Palazzi}, {Perley}, {Pian}, {Rol}, {Schady}, {Starling},
  {Tanvir}, {Watson}, {Xu}, {Augusteijn}, {Grundahl}, {Telting}, \&
  {Quirion}}]{2009ApJS..185..526F}
{Fynbo}, J.~P.~U., {et~al.} 2009, \apjs, 185, 526

\bibitem[{{Galama} \& {Wijers}(2001)}]{2001ApJ...549L.209G}
{Galama}, T.~J., \& {Wijers}, R.~A.~M.~J. 2001, \apj, 549, L209

\bibitem[{{Gendre} {et~al.}(2007){Gendre}, {Galli}, {Corsi}, {Klotz}, {Piro},
  {Stratta}, {Bo{\"e}r}, \& {Damerdji}}]{2007A&A...462..565G}
{Gendre}, B., {Galli}, A., {Corsi}, A., {Klotz}, A., {Piro}, L., {Stratta}, G.,
  {Bo{\"e}r}, M., \& {Damerdji}, Y. 2007, \aap, 462, 565

\bibitem[{{Greiner} {et~al.}(2011){Greiner}, {Kr{\"u}hler}, {Klose}, {Afonso},
  {Clemens}, {Filgas}, {Hartmann}, {K{\"u}pc{\"u} Yolda{\c s}}, {Nardini},
  {Olivares E.}, {Rau}, {Rossi}, {Schady}, \& {Updike}}]{2011A&A...526A..30G}
{Greiner}, J., {et~al.} 2011, \aap, 526, A30

\bibitem[{{Hirashita} {et~al.}(2008){Hirashita}, {Nozawa}, {Takeuchi}, \&
  {Kozasa}}]{2008MNRAS.384.1725H}
{Hirashita}, H., {Nozawa}, T., {Takeuchi}, T.~T., \& {Kozasa}, T. 2008, \mnras,
  384, 1725

\bibitem[{{Hunt} {et~al.}(2011){Hunt}, {Palazzi}, {Rossi}, {Savaglio},
  {Cresci}, {Klose}, {Micha{\l}owski}, \& {Pian}}]{2011ApJ...736L..36H}
{Hunt}, L., {Palazzi}, E., {Rossi}, A., {Savaglio}, S., {Cresci}, G., {Klose},
  S., {Micha{\l}owski}, M., \& {Pian}, E. 2011, \apjl, 736, L36

\bibitem[{{Jakobsson} {et~al.}(2004){Jakobsson}, {Hjorth}, {Fynbo}, {Watson},
  {Pedersen}, {Bj{\" o}rnsson}, \& {Gorosabel}}]{2004ApJ...617L..21J}
{Jakobsson}, P., {Hjorth}, J., {Fynbo}, J.~P.~U., {Watson}, D., {Pedersen}, K.,
  {Bj{\" o}rnsson}, G., \& {Gorosabel}, J. 2004, \apj, 617, L21

\bibitem[{{Jakobsson} {et~al.}(2006){Jakobsson}, {Levan}, {Fynbo}, {Priddey},
  {Hjorth}, {Tanvir}, {Watson}, {Jensen}, {Sollerman}, {Natarajan},
  {Gorosabel}, {Castro Cer{\'o}n}, {Pedersen}, {Pursimo}, {{\'A}rnad{\'o}ttir},
  {Castro-Tirado}, {Davis}, {Deeg}, {Fiuza}, {Mykolaitis}, \&
  {Sousa}}]{2006A&A...447..897J}
{Jakobsson}, P., {et~al.} 2006, \aap, 447, 897

\bibitem[{{Jakobsson} {et~al.}(2012){Jakobsson}, {Hjorth}, {Malesani},
  {Chapman}, {Fynbo}, {Tanvir}, {Milvang-Jensen}, {Vreeswijk}, {Letawe}, \&
  {Starling}}]{2012arXiv1205.3490J}
---. 2012, arXiv/1205.3490

\bibitem[{{Kaneko} {et~al.}(2008){Kaneko}, {Gonz{\'a}lez}, {Preece}, {Dingus},
  \& {Briggs}}]{2008ApJ...677.1168K}
{Kaneko}, Y., {Gonz{\'a}lez}, M.~M., {Preece}, R.~D., {Dingus}, B.~L., \&
  {Briggs}, M.~S. 2008, \apj, 677, 1168

\bibitem[{{Kann} {et~al.}(2010){Kann}, {Klose}, {Zhang}, {Malesani}, {Nakar},
  {Pozanenko}, {Wilson}, {Butler}, {Jakobsson}, {Schulze}, {Andreev},
  {Antonelli}, {Bikmaev}, {Biryukov}, {B{\"o}ttcher}, {Burenin}, {Castro
  Cer{\'o}n}, {Castro-Tirado}, {Chincarini}, {Cobb}, {Covino}, {D'Avanzo},
  {D'Elia}, {Della Valle}, {de Ugarte Postigo}, {Efimov}, {Ferrero}, {Fugazza},
  {Fynbo}, {G{\aa}lfalk}, {Grundahl}, {Gorosabel}, {Gupta}, {Guziy}, {Hafizov},
  {Hjorth}, {Holhjem}, {Ibrahimov}, {Im}, {Israel}, {Je{\'l}inek}, {Jensen},
  {Karimov}, {Khamitov}, {Kizilo{\v g}lu}, {Klunko}, {Kub{\'a}nek}, {Kutyrev},
  {Laursen}, {Levan}, {Mannucci}, {Martin}, {Mescheryakov}, {Mirabal},
  {Norris}, {Ovaldsen}, {Paraficz}, {Pavlenko}, {Piranomonte}, {Rossi},
  {Rumyantsev}, {Salinas}, {Sergeev}, {Sharapov}, {Sollerman}, {Stecklum},
  {Stella}, {Tagliaferri}, {Tanvir}, {Telting}, {Testa}, {Updike}, {Volnova},
  {Watson}, {Wiersema}, \& {Xu}}]{2010ApJ...720.1513K}
{Kann}, D.~A., {et~al.} 2010, \apj, 720, 1513

\bibitem[{{Kr{\"u}hler} {et~al.}(2011){Kr{\"u}hler}, {Greiner}, {Schady},
  {Savaglio}, {Afonso}, {Clemens}, {Elliott}, {Filgas}, {Gruber}, {Kann},
  {Klose}, {K{\"u}pc{\"u}-Yolda{\c s}}, {McBreen}, {Olivares}, {Pierini},
  {Rau}, {Rossi}, {Nardini}, {Nicuesa Guelbenzu}, {Sudilovsky}, \&
  {Updike}}]{2011A&A...534A.108K}
{Kr{\"u}hler}, T., {et~al.} 2011, \aap, 534, A108

\bibitem[{{Kr{\"u}hler} {et~al.}(2012){Kr{\"u}hler}, {Malesani},
  {Milvang-Jensen}, {Fynbo}, {Hjorth}, {Jakobsson}, {Levan}, {Sparre},
  {Tanvir}, \& {Watson}}]{2012arXiv1205.4036K}
---. 2012, arXiv/1205.4036

\bibitem[{{Maiolino} {et~al.}(2004){Maiolino}, {Schneider}, {Oliva}, {Bianchi},
  {Ferrara}, {Mannucci}, {Pedani}, \& {Roca Sogorb}}]{2004Natur.431..533M}
{Maiolino}, R., {Schneider}, R., {Oliva}, E., {Bianchi}, S., {Ferrara}, A.,
  {Mannucci}, F., {Pedani}, M., \& {Roca Sogorb}, M. 2004, \nat, 431, 533

\bibitem[{{Mateos} {et~al.}(2010){Mateos}, {Carrera}, {Page}, {Watson},
  {Corral}, {Tedds}, {Ebrero}, {Krumpe}, {Schwope}, \&
  {Ceballos}}]{2010A&A...510A..35M}
{Mateos}, S., {et~al.} 2010, \aap, 510, A35

\bibitem[{{Matsuura} {et~al.}(2011){Matsuura}, {Dwek}, {Meixner}, {Otsuka},
  {Babler}, {Barlow}, {Roman-Duval}, {Engelbracht}, {Sandstrom},
  {Laki{\'c}evi{\'c}}, {van Loon}, {Sonneborn}, {Clayton}, {Long}, {Lundqvist},
  {Nozawa}, {Gordon}, {Hony}, {Panuzzo}, {Okumura}, {Misselt}, {Montiel}, \&
  {Sauvage}}]{2011Sci...333.1258M}
{Matsuura}, M., {et~al.} 2011, Science, 333, 1258

\bibitem[{{M{\'e}nard} \& {Chelouche}(2009)}]{2009MNRAS.393..808M}
{M{\'e}nard}, B., \& {Chelouche}, D. 2009, \mnras, 393, 808

\bibitem[{{Micha{\l}owski} {et~al.}(2010){Micha{\l}owski}, {Watson}, \&
  {Hjorth}}]{2010ApJ...712..942M}
{Micha{\l}owski}, M.~J., {Watson}, D., \& {Hjorth}, J. 2010, \apj, 712, 942

\bibitem[{{Page} {et~al.}(2005){Page}, {Reeves}, {O'Brien}, \&
  {Turner}}]{2005MNRAS.364..195P}
{Page}, K.~L., {Reeves}, J.~N., {O'Brien}, P.~T., \& {Turner}, M.~J.~L. 2005,
  \mnras, 364, 195

\bibitem[{{Perlman} {et~al.}(2005){Perlman}, {Madejski}, {Georganopoulos},
  {Andersson}, {Daugherty}, {Krolik}, {Rector}, {Stocke}, {Koratkar}, {Wagner},
  {Aller}, {Aller}, \& {Allen}}]{2005ApJ...625..727P}
{Perlman}, E.~S., {et~al.} 2005, \apj, 625, 727

\bibitem[{{Prochaska} {et~al.}(2003){Prochaska}, {Gawiser}, {Wolfe}, {Castro},
  \& {Djorgovski}}]{2003ApJ...595L...9P}
{Prochaska}, J.~X., {Gawiser}, E., {Wolfe}, A.~M., {Castro}, S., \&
  {Djorgovski}, S.~G. 2003, \apj, 595, L9

\bibitem[{{Saez} {et~al.}(2011){Saez}, {Brandt}, {Shemmer}, {Chomiuk}, {Lopez},
  {Marshall}, {Miller}, \& {Vignali}}]{2011ApJ...738...53S}
{Saez}, C., {Brandt}, W.~N., {Shemmer}, O., {Chomiuk}, L., {Lopez}, L.~A.,
  {Marshall}, H.~L., {Miller}, B.~P., \& {Vignali}, C. 2011, \apj, 738, 53

\bibitem[{{Sambruna} {et~al.}(2007){Sambruna}, {Tavecchio}, {Ghisellini},
  {Donato}, {Holland}, {Markwardt}, {Tueller}, \&
  {Mushotzky}}]{2007ApJ...669..884S}
{Sambruna}, R.~M., {Tavecchio}, F., {Ghisellini}, G., {Donato}, D., {Holland},
  S.~T., {Markwardt}, C.~B., {Tueller}, J., \& {Mushotzky}, R.~F. 2007, \apj,
  669, 884

\bibitem[{{Sari} {et~al.}(1998){Sari}, {Piran}, \&
  {Narayan}}]{1998ApJ...497L..17S}
{Sari}, R., {Piran}, T., \& {Narayan}, R. 1998, \apj, 497, L17

\bibitem[{{Schady} {et~al.}(2011){Schady}, {Savaglio}, {Kr{\"u}hler},
  {Greiner}, \& {Rau}}]{2011A&A...525A.113S}
{Schady}, P., {Savaglio}, S., {Kr{\"u}hler}, T., {Greiner}, J., \& {Rau}, A.
  2011, \aap, 525, A113

\bibitem[{{Schady} {et~al.}(2010){Schady}, {Page}, {Oates}, {Still}, {de
  Pasquale}, {Dwelly}, {Kuin}, {Holland}, {Marshall}, \&
  {Roming}}]{2010MNRAS.401.2773S}
{Schady}, P., {et~al.} 2010, \mnras, 401, 2773

\bibitem[{{Schlegel} {et~al.}(1998){Schlegel}, {Finkbeiner}, \&
  {Davis}}]{1998ApJ...500..525S}
{Schlegel}, D.~J., {Finkbeiner}, D.~P., \& {Davis}, M. 1998, \apj, 500, 525

\bibitem[{{Starling} {et~al.}(2005){Starling}, {Vreeswijk}, {Ellison}, {Rol},
  {Wiersema}, {Levan}, {Tanvir}, {Wijers}, {Tadhunter}, {Zaurin}, {Gonzalez
  Delgado}, \& {Kouveliotou}}]{2005A&A...442L..21S}
{Starling}, R.~L.~C., {et~al.} 2005, \aap, 442, L21

\bibitem[{{Svensson} {et~al.}(2011){Svensson}, {Tanvir}, {Perley},
  {Michalowski}, {Page}, {Bloom}, {Cenko}, {Hjorth}, {Jakobsson}, {Watson}, \&
  {Wheatley}}]{2011arXiv1109.3167S}
{Svensson}, K.~M., {et~al.} 2011, arXiv/1109.3167

\bibitem[{{Tavecchio} {et~al.}(2007){Tavecchio}, {Maraschi}, {Ghisellini},
  {Kataoka}, {Foschini}, {Sambruna}, \& {Tagliaferri}}]{2007ApJ...665..980T}
{Tavecchio}, F., {Maraschi}, L., {Ghisellini}, G., {Kataoka}, J., {Foschini},
  L., {Sambruna}, R.~M., \& {Tagliaferri}, G. 2007, \apj, 665, 980

\bibitem[{{van der Horst} {et~al.}(2009){van der Horst}, {Kouveliotou},
  {Gehrels}, {Rol}, {Wijers}, {Cannizzo}, {Racusin}, \&
  {Burrows}}]{2009ApJ...699.1087V}
{van der Horst}, A.~J., {Kouveliotou}, C., {Gehrels}, N., {Rol}, E., {Wijers},
  R.~A.~M.~J., {Cannizzo}, J.~K., {Racusin}, J., \& {Burrows}, D.~N. 2009,
  \apj, 699, 1087

\bibitem[{{Vergani} {et~al.}(2009){Vergani}, {Petitjean}, {Ledoux},
  {Vreeswijk}, {Smette}, \& {Meurs}}]{2009A&A...503..771V}
{Vergani}, S.~D., {Petitjean}, P., {Ledoux}, C., {Vreeswijk}, P., {Smette}, A.,
  \& {Meurs}, E.~J.~A. 2009, \aap, 503, 771

\bibitem[{{Vladilo}(1998)}]{1998ApJ...493..583V}
{Vladilo}, G. 1998, \apj, 493, 583

\bibitem[{{Vreeswijk} {et~al.}(2006){Vreeswijk}, {Smette}, {Fruchter},
  {Palazzi}, {Rol}, {Wijers}, {Kouveliotou}, {Kaper}, {Pian}, {Masetti},
  {Frontera}, {Hjorth}, {Gorosabel}, {Piro}, {Fynbo}, {Jakobsson}, {Watson},
  {O'Brien}, \& {Ledoux}}]{2006A&A...447..145V}
{Vreeswijk}, P.~M., {et~al.} 2006, \aap, 447, 145

\bibitem[{{Watson}(2011)}]{2011A&A...533A..16W}
{Watson}, D. 2011, \aap, 533, A16

\bibitem[{{Watson} {et~al.}(2007){Watson}, {Hjorth}, {Fynbo}, {Jakobsson},
  {Foley}, {Sollerman}, \& {Wijers}}]{2007ApJ...660L.101W}
{Watson}, D., {Hjorth}, J., {Fynbo}, J.~P.~U., {Jakobsson}, P., {Foley}, S.,
  {Sollerman}, J., \& {Wijers}, R.~A.~M.~J. 2007, \apjl, 660, L101

\bibitem[{{Watson} {et~al.}(2004){Watson}, {McBreen}, {Hanlon}, {Reeves},
  {Smith}, {Perlman}, {Stocke}, \& {Rector}}]{2004A&A...418..459W}
{Watson}, D., {McBreen}, B., {Hanlon}, L., {Reeves}, J.~N., {Smith}, N.,
  {Perlman}, E., {Stocke}, J., \& {Rector}, T.~A. 2004, \aap, 418, 459

\bibitem[{{Watson} {et~al.}(2002){Watson}, {Reeves}, {Osborne}, {Tedds},
  {O'Brien}, {Tomas}, \& {Ehle}}]{2002A&A...395L..41W}
{Watson}, D., {Reeves}, J.~N., {Osborne}, J.~P., {Tedds}, J.~A., {O'Brien},
  P.~T., {Tomas}, L., \& {Ehle}, M. 2002, \aap, 395, L41

\bibitem[{{Wilms} {et~al.}(2000){Wilms}, {Allen}, \&
  {McCray}}]{2000ApJ...542..914W}
{Wilms}, J., {Allen}, A., \& {McCray}, R. 2000, \apj, 542, 914

\bibitem[{{Yuan} {et~al.}(2006){Yuan}, {Fabian}, {Worsley}, \&
  {McMahon}}]{2006MNRAS.368..985Y}
{Yuan}, W., {Fabian}, A.~C., {Worsley}, M.~A., \& {McMahon}, R.~G. 2006,
  \mnras, 368, 985

\bibitem[{{Zafar} {et~al.}(2011{\natexlab{a}}){Zafar}, {Watson}, {Fynbo},
  {Malesani}, {Jakobsson}, \& {de Ugarte Postigo}}]{2011A&A...532A.143Z}
{Zafar}, T., {Watson}, D., {Fynbo}, J.~P.~U., {Malesani}, D., {Jakobsson}, P.,
  \& {de Ugarte Postigo}, A. 2011{\natexlab{a}}, \aap, 532, A143

\bibitem[{{Zafar} {et~al.}(2011{\natexlab{b}}){Zafar}, {Watson}, {Tanvir},
  {Fynbo}, {Starling}, \& {Levan}}]{2011ApJ...735....2Z}
{Zafar}, T., {Watson}, D.~J., {Tanvir}, N.~R., {Fynbo}, J.~P.~U., {Starling},
  R.~L.~C., \& {Levan}, A.~J. 2011{\natexlab{b}}, \apj, 735, 2

\end{thebibliography}
\end{document}